\documentclass[aps,prc,showpacs,showkeys,superscriptaddress,nofootinbib,twocolumn,floatfix]{revtex4}

\usepackage{bm}
\usepackage{placeins}
\usepackage{graphicx}
\usepackage{amsfonts}
\usepackage{bbm}
\usepackage{multirow}
\usepackage{graphicx,color,amsmath,amssymb}

\usepackage{hyperref}
\hypersetup{
            colorlinks=true,
            linkcolor=blue,
            anchorcolor=blue,
            citecolor=blue}

\def\eg{{\em e.g.}}
\def\ie{{\em i.e.}}

\newcommand{\beq}{\begin{equation}}
\newcommand{\eeq}{\end{equation}}
\newcommand{\bea}{\begin{eqnarray}}
\newcommand{\eea}{\end{eqnarray}}

\newcommand{\gtsim}{\raisebox{-4pt}{$\,\stackrel{\textstyle >}{\sim}\,$}}

\begin{document}

\title{$T$-matrix Analysis of Static Wilson Line Correlators from Lattice QCD at Finite Temperature}

\author{Zhanduo Tang}
\affiliation{Cyclotron Institute and Department of Physics and Astronomy,
Texas A\&M University, College Station, TX
  77843-3366, USA}
  \author{Swagato Mukherjee}  \author{Peter Petreczky}
  \affiliation{Physics Department, Brookhaven National Laboratory, Upton, New York 11973, USA}
\author{Ralf Rapp}
\affiliation{Cyclotron Institute and Department of Physics and Astronomy,
Texas A\&M University, College Station, TX
  77843-3366, USA}
\date{\today}

\begin{abstract}
We utilize a previously constructed thermodynamic $T$-matrix approach to the quark-gluon plasma (QGP) to calculate Wilson line correlators (WLCs) of a static quark-antiquark pair and apply them to the results from 2+1-flavor lattice-QCD (lQCD) computations with realistic pion mass. The self-consistent $T$-matrix results, which include constraints from the lQCD equation of state in the light-parton sector, can describe the lQCD data for WLCs fairly well once refinements of the input parameters are implemented. In particular, the input potential requires less screening than used in previous $T$-matrix analyses. 
Pertinent predictions for the spectral and transport properties of the QGP are discussed, including the spatial diffusion coefficient for heavy quarks which turns out to have a rather weak temperature dependence, in approximate agreement with recent lQCD results.
\end{abstract}

\maketitle


\section{Introduction} 
\label{sec_intro} 
The study of the quark-gluon plasma (QGP) provides unique opportunities to understand how emergent (many-body) phenomena arise from the fundamental interactions between the partons in the Quantum Chromodynamics (QCD). In particular, heavy-flavor (HF) particles are widely regarded as excellent probes of the transport and hadronization properties of the QCD medium in ultrarelativistic heavy-ion collisions (URHICs)~\cite{Prino:2016cni,Rapp:2018qla,Dong:2019unq}, for several reasons.
Heavy-quark and -antiquark pairs are produced in initial hard processes and approximately conserved throughout the evolution of fireball. Their subsequent propagation through the medium can be characterized by well-defined transport coefficients, most notably the spatial diffusion coefficient. The interactions between heavy quarks and the medium occur with small energy transfer and thus enable the use of potential approximations. The spectra of HF particles can preserve a memory of their interaction history as their thermalization time, which is parametrically enhanced by their mass-over-temperature ratio, is comparable to or (for bottom) even longer than the fireball lifetime.

Bound states of heavy quark and antiquark, \ie, quarkonia, provide further insights into the properties of the QGP in URHICs~\cite{Rapp:2008tf,Kluberg:2009wc,Braun-Munzinger:2009dzl}, 
as their in-medium properties are closely related to the in-medium QCD force. These are, in turn, critical inputs to their transport properties which govern the abundance and transverse-momentum spectra of quarkonia~\cite{Du:2019tjf}. 
For example, quarkonium kinetics critically depends on their inelastic reaction rates, which, in turn, depend on the in-medium binding energies and are closely related to individual heavy-quark (HQ) interactions with the medium. Lattice-QCD  (lQCD) computations have provided ample information on the in-medium properties of heavy quarkonia through HQ free energies and Euclidean correlators
\cite{Karsch:2002wv,Datta:2003ww,Petreczky:2004pz,Mocsy:2004bv,Kaczmarek:2005ui,Jakovac:2006sf,Aarts:2007pk,Petreczky:2008px,Aarts:2010ek,Aarts:2011sm,Ding:2012sp,Kim:2014iga,Bazavov:2014cta,Bazavov:2018wmo,Larsen:2019bwy,Larsen:2019zqv,Larsen:2020rjk,Petreczky:2021zmz}
which constrain the spectral functions that can be computed in effective models and that subsequently serve as an input to phenomenological applications~\cite{Rapp:2009my,Mocsy:2013syh,He:2022ywp}.    

In the present paper we focus on for Wilson line correlators (WLCs) of a static quark antiquark pair at finite temperature. Pertinent lQCD results have recently been obtained using realistic 2+1-flavor lQCD calculations~\cite{Bala:2021fkm}. When compared to predictions from hard-thermal loop (HTL) perturbation theory, marked disagreement was found. On the other hand, when using schematic spectral functions based on parametrizations using different ans\"atze, it was inferred that the underlying potential exhibits a relatively weak screening, even at rather large distances, while the widths of the spectral peaks turned out to be substantial, albeit quantitatively with a rather large spread. These results were refined and essentially confirmed in a more recent analysis~\cite{Bazavov:2023dci}, and call for microscopic analysis in a nonperturbative approach. 
Toward this end, we will employ 
quantum many-body theory, \ie, a thermodynamic $T$-matrix approach, that has been previously constructed to describe the interactions between partons in a strongly coupled QGP~\cite{Cabrera:2006wh,Riek:2010fk}. Utilizing a Hamiltonian  with a non-perturbative input potential, the 1- and 2-body correlation functions evaluated selfconsistently~\cite{Mannarelli:2005pz,Riek:2010py,ZhanduoTang:2023tdg}. In the vacuum, the kernel is the standard Cornell potential, while its finite-temperature corrections have been constrained by lQCD data for the HQ free energy and equation of state (EoS). The WLCs from lQCD provide a novel opportunity to improve the constraints on the in-medium potential, especially since their dependence on euclidean time, $\tau$, for different values of spatial separation, $r$, provides a much extended dynamical reach compared to the HQ free energies which are evaluated at $\tau=1/T$.

The structure of this article is as follows. In Sec.~~\ref{sec_TM}, we briefly recollect the key components of the thermodynamic $T$-matrix approach as needed in the present context. In Sec.~\ref{sec_WLC} introduces the formalism of static WLCs and how to compute them within the $T$-matrix framework. The constraints imposed on the in-medium corrections to the potential from lQCD data for EoS and static WLCs are detailed in Sec.~\ref{sec_fits}. In Sec.~\ref{sec_transport} we compute the transport coefficients of charm and static quarks predicted by the newly inferred potential and compare the diffusion coefficient to recent lQCD results. Our summary and conclusions are contained in Sec.~\ref{sec_concl}.

\section{$T$-matrix Approach}
\label{sec_TM} 
The thermodynamic $T$-matrix is a quantum many-body approach to evaluate 1- and 2-body correlation functions selfconsistently by resumming an infinite series of ladder diagrams; thus, it is suitable to study both bound and scattering states for strongly interacting systems.  Originally devised for the study of HF particles within the QGP context~\cite{Mannarelli:2005pz,Riek:2010fk}, this approach was subsequently extended to encompass the light-parton sector~\cite{Liu:2017qah}. By reducing the 4-dimensional (4D) Bethe-Salpeter equation into a more manageable 3D Lippmann-Schwinger equation~\cite{Brockmann:1996xy}, followed by a partial-wave expansion, one obtains a 1D scattering equation~\cite{Liu:2017qah,ZhanduoTang:2023tdg},
\begin{eqnarray}
\ T_{ij}^{L,a} ( z,p,p')&=&V_{ij}^{{L,a}} (p,p') +\frac{2}{\pi } \int_{-\infty}^{\infty}k^{2}dk V_{ij}^{{L,a}} (p,k)
\nonumber \\
&&\times G_{ij}^{0} (z,k)T_{ij}^{L,a} ( z,k,p') \ , 
\label{Tmat}
\end{eqnarray}
that features the intermediate 2-parton propagator, 
\begin{eqnarray}
G_{i j}^{0}(z, k)&=& \int_{-\infty}^{\infty} d \omega_{1} d \omega_{2} \frac{\left[1 \pm n_{i}\left(\omega_{1}\right) \pm n_{j}\left(\omega_{2}\right)\right]}{z-\omega_{1}-\omega_{2}}\nonumber \\
&& \quad \times \rho_{i}\left(\omega_{1},k\right) \rho_{j}\left(\omega_{2}, k\right) \ ,
\label{G2}
\end{eqnarray}
which is a convolution of two single-parton spectral functions, 
\begin{equation} 
\rho_{i}\left(\omega, k\right)=-\frac{1}{\pi} \operatorname{Im}G_{i}(\omega+i \epsilon,k) \ ,
\label{rhoi}
\end{equation}
which are given in terms of the (positive-energy projected) propagators, 
\begin{equation}
G_{i}(\omega,k) =  1/[\omega-\varepsilon_i(k) -\Sigma_i(k)]  \ .
\label{Gi}
\end{equation}
In Eq.~(\ref{Tmat}),  $V_{ij}^{L,a}$ denotes the potential between particle $i$ and $j$ in color channel $a$ with angular-momentum $L$, the $n_{i}$ are Bose ($+$) or Fermi ($-$) distribution functions, $\varepsilon_i=\sqrt{M_{i}^{2}+k^{2}}$ on-shell energies with particle mass $M_{i}=-\frac{1}{2} \int \frac{d^{3} \mathbf{p}}{(2 \pi)^{3}} V_{i  \bar{i}}^{a=1}(\mathbf{p})+M^0_{i}$ composed of the selfenergy from the color-singlet ($a=1$) potential and the bare mass, $M^0_{i}$; the $p, p'$ are the moduli of incoming and outgoing momenta in the center-of-mass (CM) frame. The 1-particle selfenergies, $\Sigma_{i}(k)$, in Eq.~(\ref{Gi}) are calculated by closing the $T$-matrix with an in-medium 1-parton propagator~\cite{Liu:2017qah,ZhanduoTang:2023tdg}.

The static in-medium potential in color-singlet state,
\begin{equation}
\widetilde{V}(r,T) = -\frac{4}{3} \alpha_{s} [\frac{e^{-m_{d} r}}{r} + m_{d}] -\frac{\sigma}{m_s} [e^{-m_{s} r-\left(c_{b} m_{s} r\right)^{2}}-1] \ ,
\end{equation}
reduces to the well-known Cornell potential, $\widetilde{V}(r) = -\frac{4}{3} \frac{\alpha_{s}}{r}  +\sigma r$, in vacuum, where the coupling constant $\alpha_s=0.27$, and string tension, $\sigma=0.225$ $\textup{GeV}^2$, are fitted to the vacuum free energy from lQCD data~\cite{Cheng:2007jq,Petreczky:2010yn,Mocsy:2013syh,Bazavov:2014kva,HotQCD:2014kol,Bazavov:2018wmo}, with a string breaking distance of $r_{SB}=1$\,fm. For the in-medium potential, $m_d$ ($m_s$) is the Debye screening mass for color-Coulomb (confining) interaction; the parameter $c_b$ in the quadratic term of the exponential for the confining term controls the saturation of the confining potential, essentially mimicking in-medium string breaking. For the interactions between particles with finite masses, the static potential in momentum space acquires relativistic corrections induced by the underlying Lorentz structure~\cite{Riek:2010fk}
\begin{equation}
\begin{aligned}
&V_{ij}\left(\mathbf{p}, \mathbf{p}^{\prime}\right)=\mathcal{R}_{ij}^{vec} V^{vec}\left(\mathbf{p}-\mathbf{p}^{\prime}\right)+\mathcal{R}_{ij}^{sca}V^{sca}\left(\mathbf{p}-\mathbf{p}^{\prime}\right) \ ,
\end{aligned}
\label{eq_V}
\end{equation}
where $V^{vec}$ ($V^{sca}$) denotes the static vector (scalar) potential. One has 
\bea
\mathcal{R}_{ij}^{vec}&=&\sqrt{1+\frac{p^{2}}{\varepsilon _{i}(p)\varepsilon_{j}(p)}}\sqrt{1+\frac{p'^{2}}{\varepsilon _{i}(p')\varepsilon _{j}(p')}},
\\
\mathcal{R}_{ij}^{sca}&=&\sqrt{\frac{M_{i}M_{j}}{\varepsilon_i(p)\varepsilon_{j}(p)}}\sqrt{\frac{M_{i}M_{j}}{\varepsilon_i(p')\varepsilon_{j}(p')}} \ .
\label{eq_R correction}
\eea
The color-Coulomb potential is characterized by an entirely vector Lorentz structure, while the confining potential is commonly assumed to be scalar~\cite{Mur:1992xv,Lucha:1991vn}, in which case $V^{vec}=V_\mathcal{C}$ and $V^{sca}=V_\mathcal{S}$. However, Refs.~\cite{Brambilla:1996aq,Brambilla:1997kz,Szczepaniak:1996tk,Ebert:1997nk,Ebert:2002pp} have suggested that the confining potential exhibits a combination of vector and scalar Lorentz structures, expressed as $V^{vec}=V_\mathcal{C}+(1-\chi)V_\mathcal{S}$ and $V^{sca}=\chi V_\mathcal{S}$, rather than being exclusively scalar. In this context, the key parameter is the mixing coefficient, $\chi$, wherein $\chi=1$ represents a purely scalar confining potential, and values below one admix a vector component. In a previous study of 1/$m_Q$ corrections ($m_Q$: HQ mass) to the HQ interaction\cite{ZhanduoTang:2023tdg} we have found that a mixing coefficient of $\chi=0.6$ leads to a marked improvement of the spin-orbit and spin-spin splittings in vacuum charmonium and bottomonium spectroscopy~\cite{ParticleDataGroup:2018ovx} over the $\chi$=1 case. In addition, the resulting charm-quark diffusion coefficient showed better agreement with 2+1-flavor lQCD data~\cite{Altenkort:2023oms}. Therefore, we also introduce the mixing effect in this study.
The potential is extended to different color channels by 
multiplication with pertinent Casimir coefficients~\cite{Liu:2017qah,ZhanduoTang:2023tdg}.

\section{Static Wilson Line Correlators from the $T$-matrix}
\label{sec_WLC} 
The static Wilson line correlator in Euclidean time, which is amenable to lQCD computations, is connected to the static $Q\bar{Q}$
spectral function $\rho_{Q\bar{Q}}$, through a Laplace transform,
\begin{equation}
W\left (r,\tau,T  \right )=\int_{-\infty}^{\infty}dE e^{-E \tau}\rho_{Q\bar{Q}}\left ( E,r,T \right ) ,
\label{wlc}
\end{equation}
where $r$ is the distance between $Q$ and $\bar{Q}$, and $E$ their total energy which is measured relative to the mass threshold of the bare HQ mass, $2M_Q^0$ (numerically taken as $2\times 10^4$ GeV). The constituent static HQ mass is the sum of the bare mass and the mass shift originating from the selfenergy, \ie, $M_Q=M_Q^0+\widetilde{V}(r\rightarrow\infty)/2$~\cite{Liu:2017qah}. The inversion of Eq.~(\ref{wlc}) is a challenging (if not ill-posed) problem, making it difficult to reconstruct spectral functions from lQCD results for WLCs. On the other hand, if one can calculate the WLCs with spectral functions obtained from an effective model using Eq.~(\ref{wlc}), the comparison of the model-derived WLCs with lattice data can test the model capabilities and provide microscopic insights.

In the $T$-matrix formalism, the $Q\bar{Q}$ spectral function takes the same form as given in Ref.~\cite{Liu:2017qah},
\begin{equation}
\rho_{Q\bar{Q}}\left ( E,r,T \right )=\frac{-1}{\pi}\mathrm{Im}\left [ \frac{1}{E-\widetilde{V}(r,T)-\Phi(r,T)\Sigma_{Q\bar{Q}}(E,T)} \right ] 
\label{rho_QQ}
\end{equation}
where $\widetilde{V}(r,T)$ is the static in-medium potential introduced in Sec.~\ref{sec_TM}.  The two-body selfenergy, 
$\Sigma_{Q\bar{Q}}$, is related to the two-body propagator by
\begin{equation}
\left[G_{Q \bar{Q}}^0(E)\right]^{-1}=E-\widetilde{V}(r\rightarrow\infty)-\Sigma_{Q \bar{Q}}(E) \ .
\label{eq_Sigma_QQ}
\end{equation}
The two-body selfenergy is reduced by interference effects that depend on the relative distance between the quarks and are sometimes referred to the ``imaginary part of potential"~\cite{Laine:2006ns}. In the $T$-matrix formalism, these correspond to 3-body diagrams, which are difficult to calculate explicitly~\cite{Liu:2017qah}. However, one can approximately implement them through an $r$-dependent suppression factor (or interference function), $\phi(r)$~\cite{Liu:2017qah}, \ie, $\Sigma_{Q \bar{Q}}(E,r)=\Sigma_{Q \bar{Q}}(E)\phi(r)$. In perturbation theory, $\phi(r)$ has been calculated, essentially corresponding to an atanh-function (which vanishes for $r\to0$ in the color-singlet channel and goes to 1 for $r\to \infty$)~\cite{Laine:2006ns}. In a nonperturbative setting, it has been supplemented with an extra stretch-factor in the argument, which has been determined based on lQCD constraints from the static free energies~\cite{Liu:2017qah}. 
The interference effect is expected to be significant for deeply bound heavy quarkonia
and has been found to improve the description of Euclidean quarkonium correlators within the $T$-matrix approach~\cite{Liu:2017qah}.

To facilitate the understanding of the physical meaning of the lattice results for the WLCs and to what extent these can constrain the $Q\bar{Q}$ spectral function, Ref.~\cite{Bala:2021fkm} has defined its $n^{\rm th}$-order cumulants as
\begin{eqnarray}
m_1(r, \tau, T)&=&-\partial_\tau \ln W(r, \tau, T) \ ,
\label{m1}
\\
m_n&=&\partial_\tau m_{n-1}(r, \tau, T), n>1 \ .
\end{eqnarray}
The first cumulant, $m_1$, essentially corresponds to an effective mass that is commonly used in lattice QCD. 
In general, the WLC contains many states at $T=0$. At large $\tau$ (\ie, small energies) it will be dominated by the ground state that defines the potential at zero temperature. At higher energies, excited states contributing to the WLC are related to hybrid potentials (such as those between $D$-mesons) which will cause $m_1$ to have a rather complex behavior at small $\tau$. At non-zero temperature, where only data at relatively small $\tau$ are available, this fact renders the analysis quite involved. We therefore employ the lQCD results for $m_1$ where the excited-state contributions to the WLC have been subtracted, following the procedure outlined in Ref.~\cite{Bala:2021fkm}, which renders a much simpler behavior at small $\tau$.

Returning to the $T$-matrix expression, Eq.~(\ref{rho_QQ}), one can show by expanding $m_1$ for $\tau\to 0$ and considering the analytic properties of the two-body selfenergy, that $m_1(r, \tau=0, T)$ reduces to the potential $\widetilde{V}(r,T)$, see Appendix~\ref{sec_Wm1} for more details. The second cumulant $m_2$, or equivalently the slope of $m_1$, is a measure of the width of $\rho_{Q\bar{Q}}$, characterizing the interacting strength between $Q\bar{Q}$ and medium.


\section{In-Medium Potentials Constrained by lQCD}
\label{sec_fits} 
In this section, we lay out the procedure of inferring the modifications to the in-medium potential by means of HQ WLCs and QGP EoS within a selfconsistent quantum many-body approach~\cite{Liu:2017qah,ZhanduoTang:2023tdg}. 
It consists of 2 self-consistency loops, as follows. Initially, the parameterization of the in-medium potential is refined by calculating the cumulants of static WLCs ($m_1$) using the $T$-matrix approach and fitting them to the lQCD data. The fit parameters in this step are the screening masses, $m_d$ (color-Coulomb) and $m_s$ (confining), as well as $c_b$ which controls the ``string breaking". Subsequently, these refined in-medium potentials are used to compute the EoS of QGP, and the main parameters in this step include the in-medium light-quark and gluon masses. Since the calculation of EoS requires the one-body spectral functions and two-body scattering amplitudes (which depend on each other through the parton selfenergies, cf.~Eqs.~(\ref{Tmat})-(\ref{Gi})), 
one has to solve a selfconsistency problem which is done by numerical iteration. The selfenergies of the heavy quarks, calculated from heavy-light $T$-matrices closed off with thermal parton spectral functions, are then reinserted into the calculation of the WLCs via the in-medium HQ spectral functions.
A more refined in-medium two-body potential is created by refitting the screening masses to the lQCD WLC data, and this potential is then processed to provide a new fit to the EoS, and the process is iterated until convergence has been achieved.

In the remainder of this section, we discuss our numerical fits of the EoS (following Ref.~\cite{Liu:2017qah}) in Sec.~\ref{ssec_EOS} and the results for the WLCs in Sec.~\ref{ssec_WLC}.

\subsection{Equation of State}
\label{ssec_EOS}
 \begin{figure*}[htbp]
\begin{minipage}[b]{1.0\linewidth}
\centering
\includegraphics[width=0.9\textwidth]{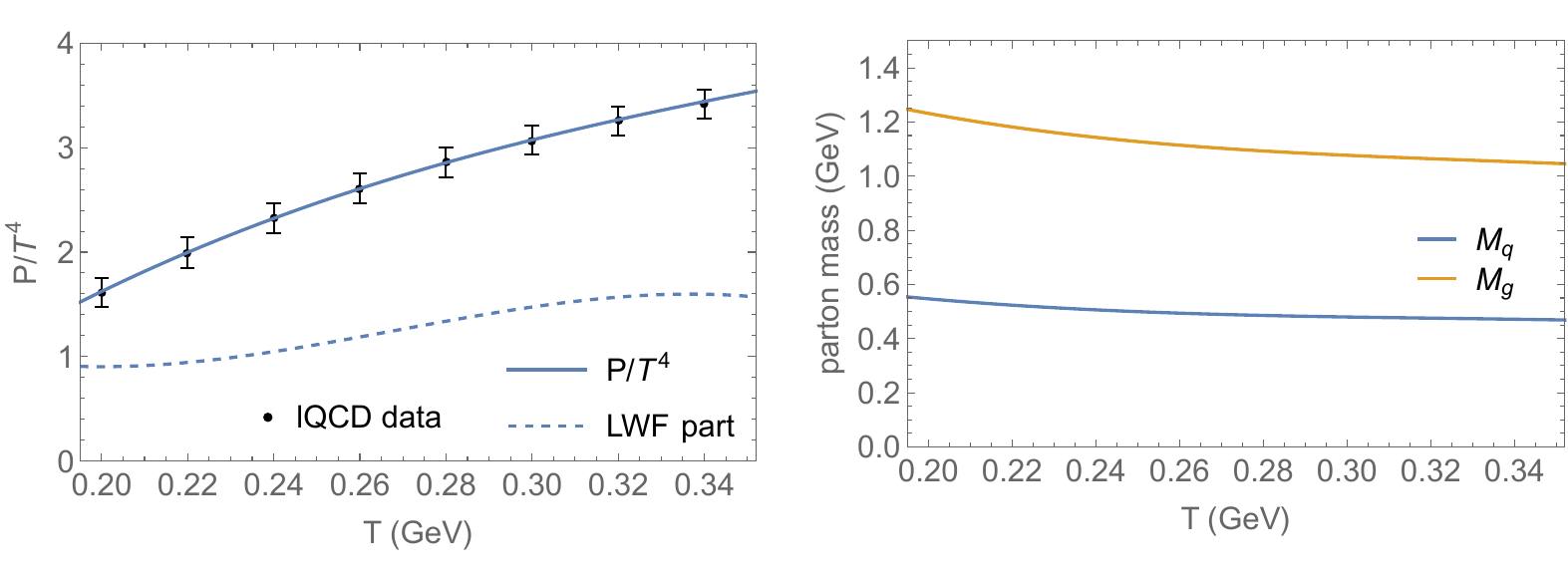}
\end{minipage}
\caption{Left panel: the pressure scaled by $T^4$ (blue line) compared to the pertinent 2+1-flavor lQCD data~\cite{HotQCD:2014kol}, as well as the LWF contribution (dashed line). Right panel: the total in-medium masses (\ie, bare plus selfenergies) for light quarks (blue) and gluons (orange) as a function of temperature.} 
\label{fig_P}
\end{figure*}
The equation of state is encoded in the pressure, $P(T,\mu)$, of a many-body system as a function of temperature and chemical potential(s). In the present context, lQCD data for the EoS are used to adjust the bare light-parton masses which are due to effects that are not explicitly part of the many-body calculations, \eg, due to quark or gluon condensates (similar to what is done in quasiparticle models). In the grand canonical ensemble the pressure is related to the grand potential per unit volume via $\Omega=-P$. 
For an interacting system, it can be calculated diagrammatically within the Luttinger-Ward-Baym (LWB) formalism where all closed-loop ``skeleton" diagrams are computed with fully dressed propagators~\cite{Luttinger:1960ua,Baym:1961zz,Baym:1962sx}, also referred to as a 2-particle irreducible approach. 
This constitutes a conserving (\ie, thermodynamically consistent) approximation scheme, which is compatible with the ladder resummation in the selfconsistent $T$-matrix and selfenergies. 
For a Hamiltonian approach to the QCD, this has been carried out in Refs.~\cite{Liu:2016ysz,Liu:2017qah}. In particular, the Luttinger-Ward functional, $\Phi$, could be resummed utilizing a matrix-logarithm resummation technique which accounts for the possibility of dynamically formed bound states contributing to the EoS. In compact form the result can be written as
\begin{equation}
\begin{aligned}
\Omega=& \sum_{j} \mp d_{j} \int d \tilde{p}\left\{\ln \left(-G_{j}(\tilde{p})^{-1}\right)\right.\\
&\left.+\left[\Sigma_{j}(\tilde{p})-\frac{1}{2} \log \Sigma_{j}(\tilde{p})\right] G_{j}(\tilde{p})\right\},
\end{aligned}
\label{eq_EOS}
\end{equation}
using the notation $\int d \tilde{p} \equiv-\beta^{-1} \sum_{n} \int d^{3} \mathbf{p} /(2 \pi)^{3}$ with $\tilde{p} \equiv\left(i \omega_{n}, \mathbf{p}\right)$ and $\beta=1/T$. The summation in Eq.~(\ref{eq_EOS}) includes all light-parton channels with the spin-color degeneracy $d_j$, and $\mp$ denotes bosons (upper) or fermions (lower). The three pieces in Eq.~(\ref{eq_EOS}), $\ln (-G^{-1})$, $\Sigma G$, and $\log \Sigma G$, correspond to the contributions from quasiparticles, selfenergies and the Luttinger-Ward functional (LWF) characterizing two-body interactions, respectively.  

We present the fits to the pressure as obtained from lQCD~\cite{HotQCD:2014kol} together with LWF contributions, as well as the corresponding light-parton masses in Fig.~\ref{fig_P}. In accordance with the findings in Refs.~\cite{Liu:2017qah,ZhanduoTang:2023tdg}, the  two-body interaction (LWF) plays an increasingly important role as temperature decreases, becoming the leading contribution at $T=0.195$\,GeV, indicating a transition in the degrees of freedom (from QGP to mesons and diquarks) in the system.

\subsection{Static Wilson Line Correlators}
\label{ssec_WLC}
\begin{figure*}[tbp]
\begin{minipage}[b]{1.0\linewidth}
\centering
\includegraphics[width=0.97\textwidth]{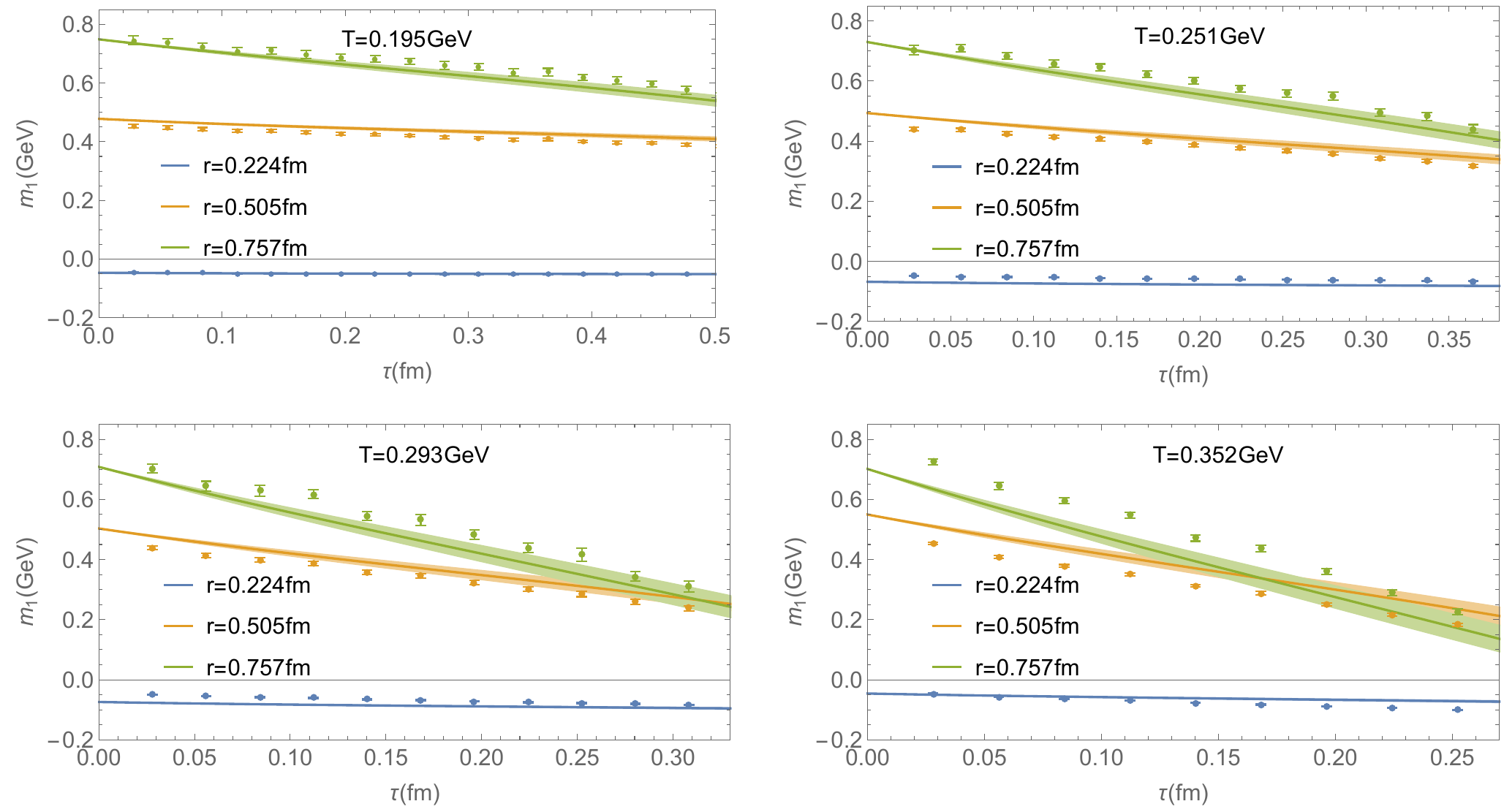}
\end{minipage}
\caption{The first cumulants of WLCs (lines) as a function of imaginary time at different temperatures and distances in comparison with the corresponding 2+1-flavor lQCD data~\cite{Bala:2021fkm}.} 
\label{fig_m1}
\end{figure*}

Next we turn to the fits to the cumulants of the WLCs, which are, in fact, part of the combined fit procedure as outlined at the beginning of this section, cf.~Fig.~\ref{fig_m1}.
A fair overall agreement with the lQCD results can be achieved for both the intersects at $\tau=0$ and the slopes of $m_1$, although the description somewhat worsens at the highest temperature. We will return to the reason for that below.
%
 \begin{figure}[htbp]
\begin{minipage}[b]{1.0\linewidth}
\centering
\includegraphics[width=0.9\textwidth]{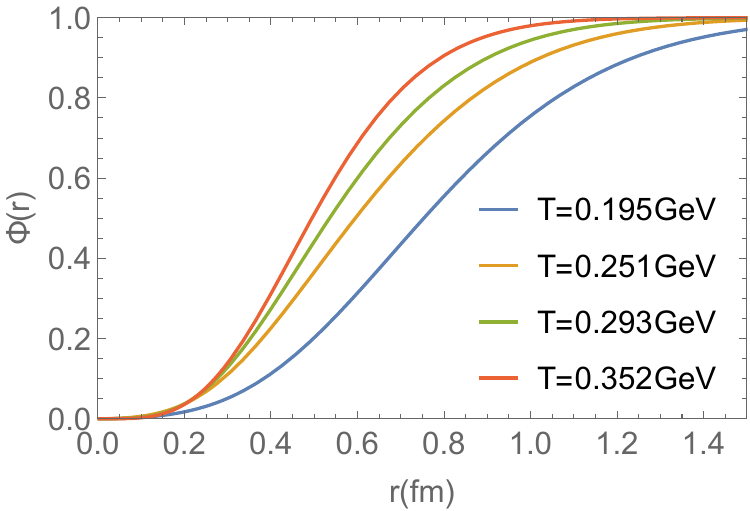}
\end{minipage}
\caption{The interference function, $\phi(r)$, as a function of distance at different temperatures as interpolated from Ref.~\cite{Liu:2017qah}.} 
\label{fig_phi}
\end{figure}
For the interference functions, $\phi(r,T)$, we simply took the results from previous Ref.~\cite{Liu:2017qah} (interpolated to the set of temperatures used here) which were based on fits to the lQCD data for HQ free energies, see Fig.~\ref{fig_phi}, but allowing for $\pm10\%$ variations to illustrate underlying uncertainties.
They show the qualitatively expected features of pushing the interference effects deeper into the bound state as temperature increases.

\begin{figure}[htbp]
\begin{minipage}[b]{1.0\linewidth}
\centering
\includegraphics[width=0.88\textwidth]{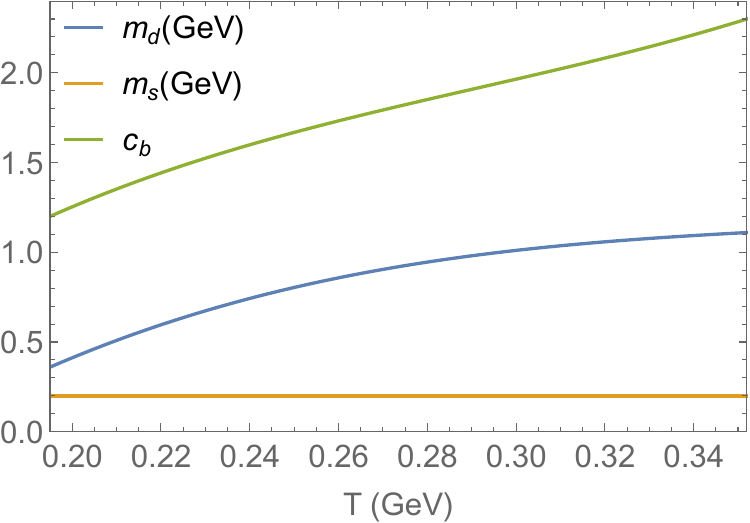}
\end{minipage}
\caption{The in-medium potential parameters as a function of temperature.} 
\label{fig_Vpara}
\end{figure}

The main difference relative to the previous $T$-matrix results lies in the screening parameters of the potential parameters as a function of temperature, shown in Fig.~\ref{fig_Vpara}. Most notably, it turns out the the fits to the WLCs can be carried out with a constant Debye mass for the string interaction: at $m_s$=0.2\,GeV is comparable to the low-$T$ values in previous work but does not increase with $T$. Instead, it turns out that an increase in the $c_b$ parameter is in order (which was assumed to be constant at $c_b$=1.3-1.55 in the previous works), implying that the screening of the string interaction is relegated to larger distances. On the other hand, the screening mass for the color-Coulomb interaction did not change much in either magnitude or temperature dependence. 

%
\begin{figure}[htbp]
\begin{minipage}[b]{1.0\linewidth}
\centering
\includegraphics[width=0.9\textwidth]{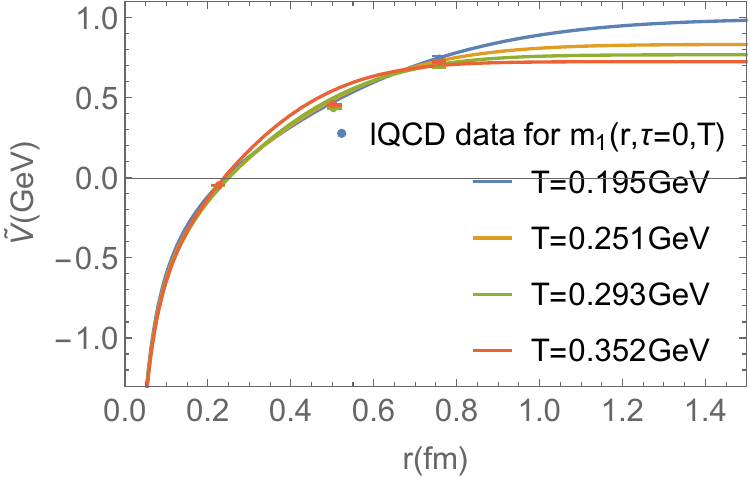}
\end{minipage}
\caption{The in-medium potentials (lines) as a function of distance at different temperatures in comparison to the 2+1-flavor lQCD results (dots) for the first cumulants of WLCs at $\tau=0$~\cite{Bala:2021fkm}.} 
\label{fig_V}
\end{figure}
The pertinent in-medium potentials are displayed in Fig.~\ref{fig_V}. Following the discussion of the screening parameters, the main difference to previous $T$-matrix extractions largely based on HQ free energies lies at the higher temperatures and larger distances where less screening is present based on the WLCs fits. We also plot lQCD data points extracted from the small-$\tau$ limit of $m_1$. They essentially overlap with the extracted potential, except at the largest temperature where the interplay of the medium and large-distance points is not quite captured, and which is at the origin of the discrepancy in the WLC fits in lower right panel in Fig.~\ref{fig_m1}.
This might be in part due to the fact that to observe string breaking on the lattice, it is not sufficient to consider WLCs or Wilson loops alone, but operators corresponding to static-light mesons should also be included in the analysis \cite{Bali:2005fu} (recall  our remark following Eq.~(\ref{m1})).
We also note that our fits utilize a mixing coefficient for the vector component of the confining potential of $\chi=0.8$, compared to 0.6 in Ref.~\cite{ZhanduoTang:2023tdg} and 1 in Ref.~\cite{Liu:2017qah}. While this choice still takes advantage of the improvement in the hyperfine splittings in the vacuum spectroscopy, it mitigates somewhat the aforementioned problem in the potential at the highest temperature which would be exacerbated for $\chi$=0.6.
%

%
\begin{figure*}[thbp]
\begin{minipage}[b]{1.0\linewidth}
\centering
\includegraphics[width=0.9\textwidth]{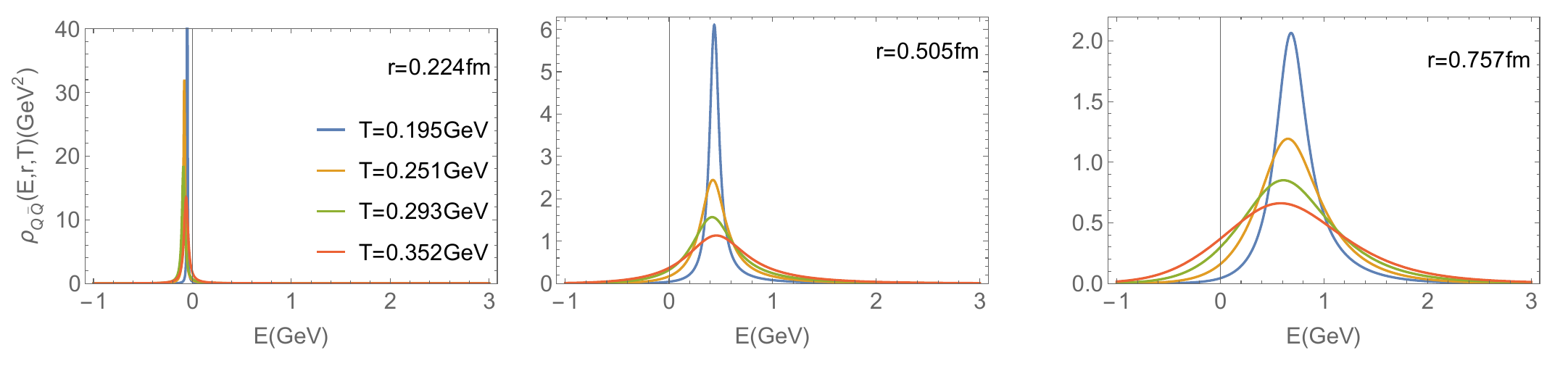}
\end{minipage}
\caption{The spectral functions (with $E$ shifted by twice static-quark mass) of static quark antiquark pair  as a function of energy at different temperatures and distances.} 
\label{fig_rhoQQ}
\end{figure*}
The bridge between the underlying potential and the WLCs is provided by the static $Q\bar{Q}$ spectral functions which are obtained from the selfconsistent solutions of the former and subsequently injected into Eq.~(\ref{wlc}) and used to compute the pertinent moment;  they are depicted in Fig.~\ref{fig_rhoQQ}. 
The effective mass of $Q\bar{Q}$ state, characterizing the pole position of $\rho_{Q\bar{Q}}$ at distance $r$, is largely determined by the potential, $\widetilde{V}(r)$ (and identical to the latter in the limit $m_1(\tau=0)$). However, they are not exactly the same since the non-zero real part of $Q\bar{Q}$ selfenergy, $\Sigma_{Q\bar{Q}}$ in Eq.~(\ref{rho_QQ}), slightly shifts the pole position away from $\widetilde{V}(r)$. Nevertheless, $m_1(\tau=0)$ remains an indicator of the $Q\bar{Q}$ effective mass, which turns out to be rather temperature-independent as a consequence of the approximately temperature-independent potentials at small and intermediate distances (as shown in Fig.~\ref{fig_V}). The increase of effective mass with increasing $r$ results from the smaller attraction (less binding) between $Q$ and $\bar{Q}$ at larger $r$.

The width of the static spectral function indicates the interaction between $Q\bar{Q}$ and the medium partons and is quantified by the imaginary part of the $Q\bar{Q}$ selfenergy, $\Phi(r,T){\rm Im}\Sigma_{Q\bar{Q}}(E,T)$, in Eq.~(\ref{rho_QQ}); its dependence on temperature and distance are consistent with the slope of $m_1$ in Fig.~\ref{fig_m1}, exhibiting a strong broadening with increasing temperature and distance. The dependence on temperature is a consequence of several competing effects: a larger QGP density and less interference tend to increase width, while a smaller interaction strength with increasing temperatures decreases it. The distance dependence has two basic components, \ie, a long-range force enables a parton to interact with a larger number of thermal partons in the heat bath, proportional to the volume of the spherical shell which grows as $r^2$~\cite{Liu:2018syc}, and the ceasing of interference effects with $\phi(r)$ approaching 1, enabling both $Q$ and $\bar Q$ to fully interact with the QGP.

\begin{figure*}[htbp]
\begin{minipage}[b]{1.0\linewidth}
\centering
\includegraphics[width=1.0\textwidth]{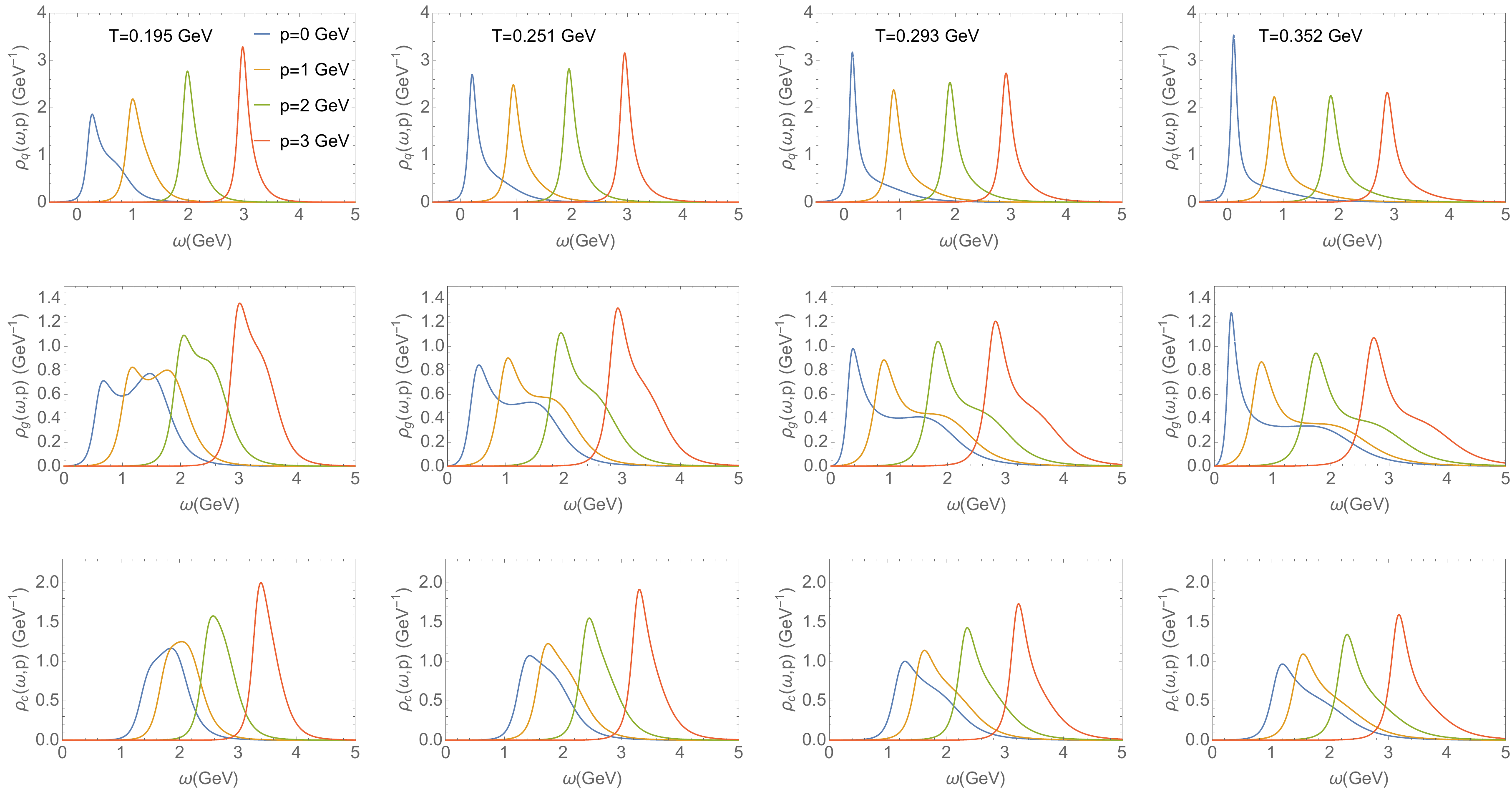}
\end{minipage}
\caption{Single-parton spectral functions for light quarks (first row), gluons (second row) and charm quarks (third row) as a function of energy for various 3-momenta in each panel. From left to right, the four columns correspond to temperatures of $T=0.195$, 0.251, 0.293 and 0.352\,GeV, respectively.} 
\label{fig_sfs}
\end{figure*}
Finally, we display in Fig.~\ref{fig_sfs} the spectral functions of light quarks, gluons and charm quarks as following from the selfconsistent $T$-matrix solution based on the WLC fit. Not surprisingly, the most significant changes are at the two higher temperatures ($T$=0.293, 352\,GeV) where the interaction strength is larger than before~\cite{ZhanduoTang:2023tdg}. In the light-parton sector this leads to broader ``quasiparticle" peaks and generates stronger collective modes on the low-energy shoulder of the ``quasiparticle peaks" at zero parton momentum; note that the effective quark mass is around 0.5\,GeV while the peaks of the collective mode are located near $\omega$$\simeq$\,0.1\,GeV. For the zero-momentum gluon spectral functions at the two higher temperatures , the collective modes emerge near $\omega$=0.2\,GeV, which is, in fact, not far from gluon condensation. It is this feature that, especially at $T$=0.352\,GeV, prevents us from improving the fit of the WLC, as a stronger input potential at large distances results in an unstable fitting procedure due to the emergence of gluon condensation (signalled by a disappearance of the low-energy peak and a real part of the propagator flipping from negative to positive). Whether this ``feature" can be turned into a framework where the bare parton masses are selfconsistently obtained from a condensed ground state is beyond the scope of this work (it would also have to be implemented at the lower temperatures). Even in the charm-quark sector the low-momentum spectral functions at the higher temperatures are significantly broader than before, again accompanied with a collective low-energy peak which implies notable deviations from a simple ``quasiparticle" spectral shape.


\section{Charm-Quark Transport Coefficients}
\label{sec_transport}
\begin{figure*}[htbp]
\begin{minipage}[b]{1.0\linewidth}
\centering
\includegraphics[width=1.0\textwidth]{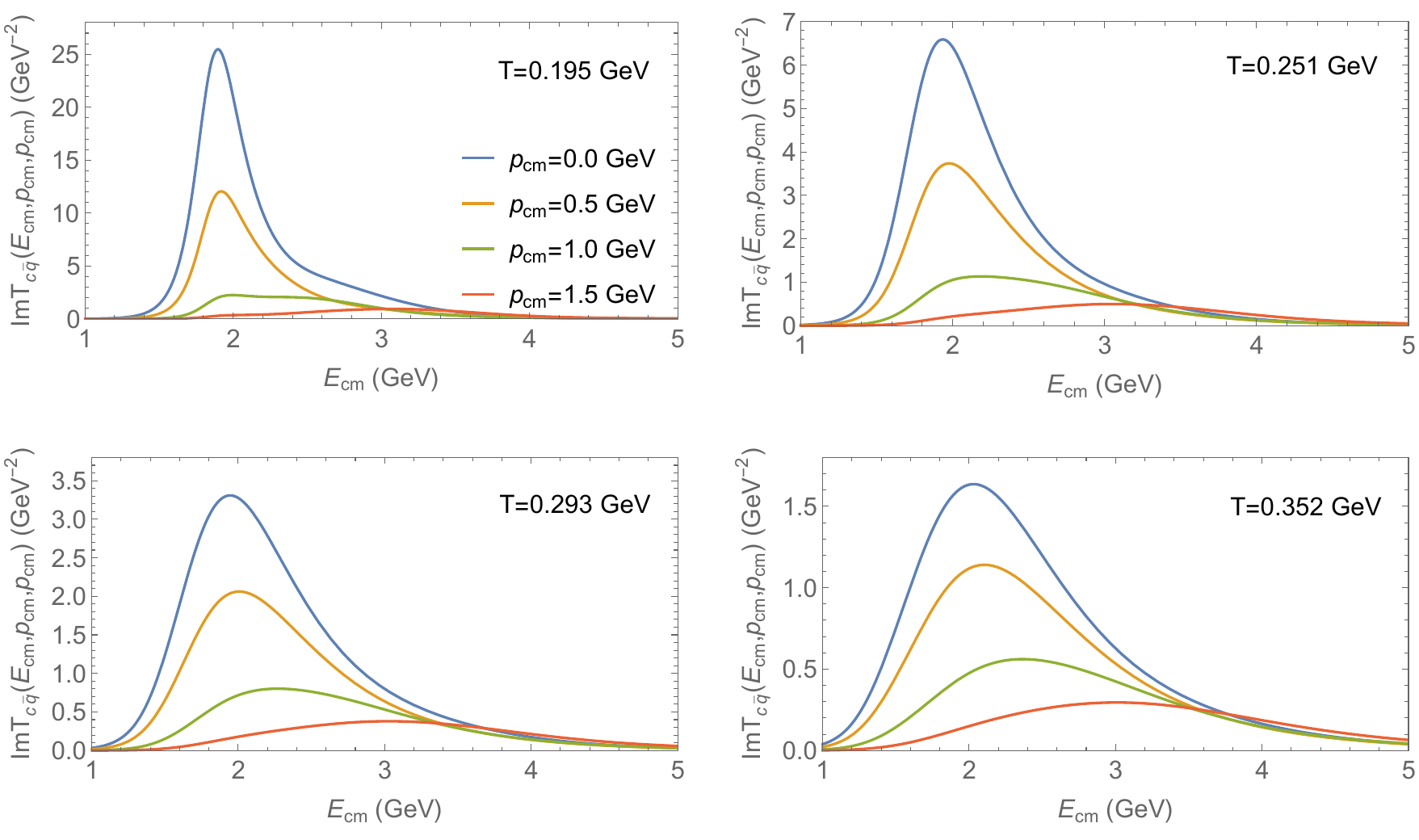}
\end{minipage}
\caption{The imaginary part of the $S$-wave charm-light $T$-matrices as a function of CM energy in the color-singlet channel at different temperatures for various CM momenta in each panel.} 
\label{fig_Tm}
\end{figure*}

We are now in position to compute the charm-quark transport properties in the QGP with the refined potential. Two main ingredients to the charm-quark transport coefficients are parton spectral functions, discussed in the previous section, and the heavy-light scattering amplitudes. The latter are displayed in Fig.~\ref{fig_Tm} for $S$-wave $c\bar{q}$ scattering in the color-singlet channel (which provides the largest contribution, and together with color-anti-triplet diquark amplitudes makes up the dominant contribution to the transport coefficient). 
In our recent work~\cite{ZhanduoTang:2023tdg}, we have found that the inclusion of a vector component in the string force (with mixing coefficient $\chi$=0.6) leads to significantly harder momentum dependence of the scattering amplitudes than the previous results for $\chi$=0~\cite{Liu:2017qah}. This trend persists here, although a little less pronounced due to the choice of $\chi$=0.8, leading to slightly reduced amplitudes for $T$=0.195\,GeV (compared to $\chi$=0.6) while they are quite comparable for $T$=0.251\,GeV at finite momentum. For the two higher temperatures, the effect of the stronger potential becomes important again.

As the resonance peaks in the heavy-light amplitudes are largely below the ``nominal" 2-parton threshold (\eg, $E_{cm}$$\simeq$2\,GeV for the $S$-wave ``$D$-meson" resonances in Fig.~\ref{fig_Tm} vs.~ 
$E_{\rm thr}$=$m_q + m_c$$\simeq$2.3\,GeV at $T$=0.195\,GeV), it is mandatory to account for the nontrivial spectral functions in the evaluation of the HQ transport coefficient. Following the previous  study~\cite{Liu:2018syc} in this context, the off-shell effects in calculating the HQ friction coefficient (relaxation rate) can be implemented by utilizing  Kadanoff-Baym equations to yield
\begin{equation}
\begin{aligned}
A(p)=& \sum_{i} \frac{1}{2 \varepsilon_{c}(p)} \int \frac{d \omega^{\prime} d^{3} \mathbf{p}^{\prime}}{(2 \pi)^{3} 2 \varepsilon_{c}\left(p^{\prime}\right)} \frac{d \nu d^{3} \mathbf{q}}{(2 \pi)^{3} 2 \varepsilon_{i}(q)} \frac{d \nu^{\prime} d^{3} \mathbf{q}^{\prime}}{(2 \pi)^{3} 2 \varepsilon_{i}\left(q^{\prime}\right)} \\
& \times \delta^{(4)} \frac{(2 \pi)^{4}}{d_{c}} \sum_{a, l, s}|M|^{2} \rho_{c}\left(\omega^{\prime}, p^{\prime}\right) \rho_{i}(\nu, q) \rho_{i}\left(\nu^{\prime}, q^{\prime}\right) \\
& \times\left[1-n_{c}\left(\omega^{\prime}\right)\right] n_{i}(\nu)\left[1 \pm n_{i}\left(\nu^{\prime}\right)\right] (1-\frac{\mathbf{p}\cdot\mathbf{p'}}{\mathbf{p}^2}) \ .
\end{aligned}
\label{eq_A(p)}
\end{equation}
Here $\delta^{(4)}$ is a short-hand notation for the energy-momentum conserving $\delta$-function in the 2$\to$2 scattering process, $d_c=6$ the spin-color degeneracy of charm quarks, and the summation, $\sum_{i}$, is over all light-flavor quarks and gluons making up the thermal medium, $i=u,\bar{u},d,\bar{d},s,\bar{s},g$ (the masses for light and strange quarks are assumed to be the same).  Only the incoming charm quark is assumed to be a quasiparticle of definite momentum and corresponding on-shell energy, while the off-shell effects are implemented by energy convolutions over the spectral functions of the light parton and the outgoing charm quark. The heavy-light scattering matrix elements, $|M|^2$ in Eq.~(\ref{eq_A(p)}), are directly related to the $T$-matrix in the CM frame which incorporates the summation over all possible two-body color and partial-wave channels~\cite{Liu:2018syc}.

\begin{figure}[htbp]
\begin{minipage}[b]{1.0\linewidth}
\centering
\includegraphics[width=1.0\textwidth]{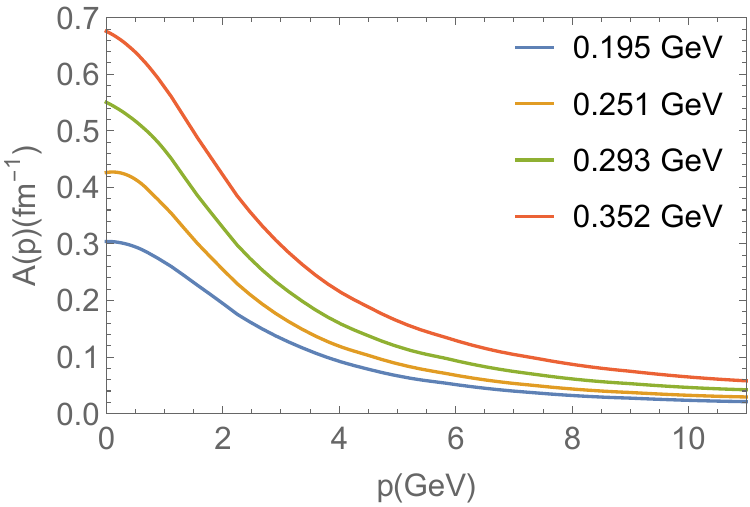}
\end{minipage}
\caption{The charm-quark friction coefficient as a function of charm-quark 3-momentum at different temperatures.} 
\label{fig_Ap}
\end{figure}
In Fig.~\ref{fig_Ap} we plot our results for the charm-quark friction coefficient, $A(p;T)$. At the lower two temperatures the WLC-based result is a little lower than the one from Ref.~\cite{ZhanduoTang:2023tdg}, mostly due to the smaller value of $\chi$ (0.8 vs.~0.6), but at temperatures $T$$\gtsim$0.3\,GeV, it becomes larger at low momentum (due to the stronger potential) and comparable at momenta $p\gtsim 8\,GeV$ (where the color-Coulomb interaction starts to dominate).

\begin{figure}[htbp]
\begin{minipage}[b]{1.0\linewidth}
\centering
\includegraphics[width=1.0\textwidth]{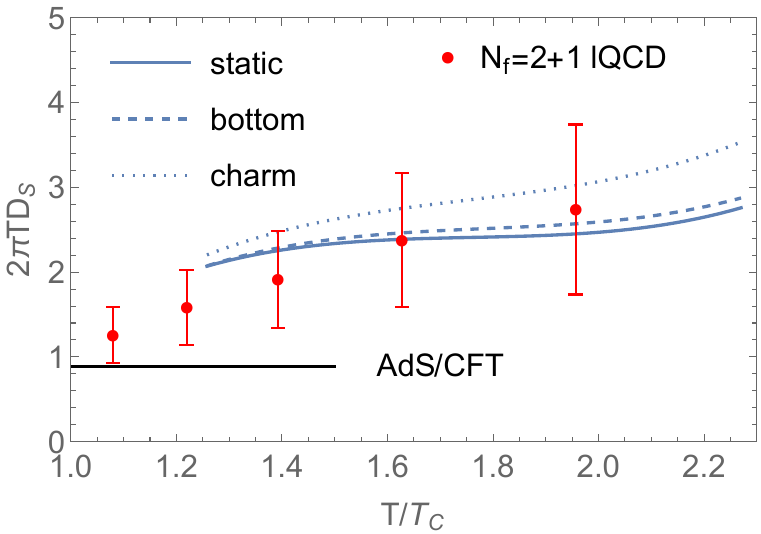}
\end{minipage}
\caption{Our result for the spatial diffusion coefficient for static (blue solid line), bottom (blue dashed line) and charm (blue dotted line) quarks as a function of temperature (scaled by $T_c$=0.155 GeV) compared to the 2+1-flavor lQCD data~\cite{Altenkort:2023oms} (red dots with error bars) and the AdS/CFT estimate~\cite{Casalderrey-Solana:2006fio} (black line). The bare quark masses for charm and bottom are 1.359 and 4.681~GeV, respectively, and we use 10\,GeV to approximate the static limit.} 
\label{fig_Ds}
\end{figure}

The spatial diffusion coefficient, $D_s=T/(M_c A(p=0))$, which is proportional to the relaxation time, $\tau_c = 1/A(p=0)$, is commonly scaled by the inverse thermal wavelength, $2\pi T$, to render a dimensionless quantity. We display the diffusion coefficient in Fig.~\ref{fig_Ds} as a function of temperature.
In the static limit our result is in fair agreement with recent lQCD data~\cite{Altenkort:2023oms}, and about a factor of 2-3 larger than the result from the AdS/CFT correspondence which is believed to provide a quantum lower bound for this quantity~\cite{Casalderrey-Solana:2006fio}. The temperature dependence is a bit weaker than our previous results, again a consequence of the stronger potential at the higher temperatures. 

The phenomenological consequences of our updated transport coefficients for HF observables in URHICs, which will also require the inclusion of radiative contributions~\cite{Liu:2020dlt}, remain to be worked out. 
%

\section{Conclusions}
\label{sec_concl}
We have applied the thermodynamic $T$-matrix approach to compute static Wilson line correlators and utlilized them to analyze pertinent lattice data. By varying the screening properties of the input potential and carrying out selfconsistent calculations of the 1- and 2-body correlation functions that encompass a description of lQCD results for the QGP equation of state, solutions were found that result in a fair agreement with the lQCD data for the first cumulants as a function of euclidean time. While the input potential at low temperatures is quite similar to previous solutions that were based on fits to heavy-quark free energies, more significant adjustments were required at temperatures above $\sim$300\,MeV, amounting to a significantly less screened (\ie, stronger) potential. This reinforces earlier findings that remnants of the confining force play a critical role in the properties of the strongly coupled QGP (well) above the critical temperature, with parton collision widths in excess of 0.5\,GeV.

An immediate consequence of the stronger potential is an enhancement of low-energy collective modes in the light-quark and gluon spectral functions that develop well below the nominal values of their masses, implying strong deviations from the quasiparticle picture. Even for charm-quark spectral functions this distortion is still significant.  We have also computed the pertinent HQ transport coefficients. Their temperature dependence turns out to be more gradual than before, and the predicted spatial diffusion coefficient in the static limit is still in approximate agreement with recent 2+1-flavor lQCD results. It will be interesting to see how these calculations fare when implemented into phenomenological applications to HF data in heavy-ion collisions.
Work in this direction is in progress. 

\acknowledgments
This work has been supported by the U.S. National Science Foundation under grant nos. PHY-1913286 and PHY-2209335, and by the The U.S. Department of Energy, Office of Science, Office of Nuclear Physics through contract No. DE-SC0012704 and the Topical Collaboration in Nuclear Theory on \textit{Heavy-Flavor Theory (HEFTY) for QCD Matter} under award no.~DE-SC0023547.

 \appendix
    
    \section{Relation between First Cumulant of WLC and Potential}
    \label{sec_Wm1}
In this Appendix we prove the identity that the first cumulant of the WLCs in the limit of vanishing euliceand time recovers the static potential,  $m_1(r, \tau=0)=\widetilde{V}(r)$ (for simplicity we suppress the temperature dependence), as mentioned at the end of Sec.~\ref{sec_WLC}. Expanding $W(r,\tau)$ in the vicinity of $\tau=0$, one obtains
\begin{eqnarray}
W(r,\tau)&=&\int_{-\infty}^{\infty}dE e^{-E \tau}\rho_{Q\bar{Q}}\left ( E,r \right ) \nonumber\\ 
&\approx&\int_{-\infty}^{\infty}dE \rho_{Q\bar{Q}}(E,r)-\tau\int_{-\infty}^{\infty}dE  E \rho_{Q\bar{Q}}(E,r)\nonumber\\
&=&1-\tau \widetilde{V}(r) \ .
\label{eq_Wm1}
\end{eqnarray}
The last identity in Eq.~(\ref{eq_Wm1}) has been proved in Ref.~\cite{Liu:2017qah} using a contour integral and the fact that the two-body selfenergy, $\Sigma_{Q\bar{Q}}$, in the spectral function, $\rho_{Q\bar{Q}}$ of Eq.~(\ref{rho_QQ}), is analytic. In addition, the identity $\int_{-\infty}^{\infty}dE \rho_{Q\bar{Q}}(E,r)=1$ is nothing but the normalization condition of spectral function. The first cumulant of the WLC, $m_1(r,\tau)$, then becomes
\begin{equation}
m_1(r, \tau)=-\partial_\tau \ln W(r, \tau)=\widetilde{V}(r),
\end{equation}
and all the higher-order terms of $\tau$ vanish at $\tau=0$.
\bibliography{refnew}

\end{document}